\documentstyle[aps,prl,multicol,epsf]{revtex}
\begin{document}
\draft
 \def\OP{\tensor P}
\def\B.#1{{\bbox{#1}}}
\def\BE {\begin{equation}}
\def\EE {\end{equation}}
\def\BEA {\begin{eqnarray}}
\def\EEA {\end{eqnarray}}
\def\Fbox#1{\vskip0ex\hbox to 8.5cm{\hfil\fboxsep0.3cm\fbox{%
  \parbox{8.0cm}{#1}}\hfil}\vskip0ex}

\title{Universal scaling exponents in shell models of turbulence: \\
Viscous effects are finite-size corrections to scaling} 
\author{Victor S. L'vov*, Itamar Procaccia and Damien Vandembroucq}
\address{Department of~~Chemical Physics, The Weizmann Institute
of Science, Rehovot 76100, Israel\\
$*$Also at the Institute of Automatization and Electrometry, 
Russian Ac. Sci., Novosibirsk 630090, Russia. } 
\maketitle 
\begin{abstract} 
In a series of recent works it was proposed that shell models of
turbulence exhibit inertial range scaling exponents that depend on the
nature of the dissipative mechanism. If true, and if one could imply a
similar phenomenon to Navier-Stokes turbulence, this finding would
cast strong doubts on the universality of scaling in turbulence.  In
this Letter we propose that these ``nonuniversalities" are just
corrections to scaling that disappear when the Reynolds number goes to
infinity.
\end{abstract}
\begin{multicols}{2}
The aim of this Letter is to question a growing consensus
\cite{SL95PRL,Kad95PF,Schorg95,Dit97PF} concerning the non-universality
of the scaling exponents that characterize correlation functions that
appear in the context of numerical studies of shell models of
turbulence. Shell models, and in particular the so-called GOY model
\cite{Gledzer,GOY}, became very popular due to their ease of
simulation and the fact that they appear to exhibit anomalous scaling
that is strongly reminiscent of the behavior of the scaling exponents
of Navier-Stokes turbulence \cite{Jensen91PRA,Piss93PFA,Benzi93PHD}.
The GOY model is a simplified reduced wavenumber analog to the
spectral Navier-Stokes equations.  The wavenumbers are represented as
shells, each of which is defined by a (real) wavenumber $k_n\equiv
k_0\lambda^n$ where $\lambda$ is the ``shell spacing". There are $N$
degrees of freedom where $N$ is the number of shells. The model
specifies the dynamics of the ``velocity" $u_n$ which is considered a
complex number, $n=1,\dots,N$.  The GOY model reads
\begin{eqnarray}
\Big[{d\over dt}&+&\nu k^{2\alpha}_n\Big]u_n
=f_n+i\big[k_{n+1}u_{n+1}u_{n+2}\nonumber\\
&-&{k_n\over 2}u_{n+1}u_{n-1}-{k_{n-1}
\over 2}u_{n-1}u_{n-2}\big]^* \ . \label{goy}
\end{eqnarray}
In this equation $^*$ stands for complex conjugation, $\nu$ is the
``viscosity", and to motivate the discussion we already specified a
one parameter family of models in which the exponent $\alpha$
determines the type of viscous dissipation.  In the following we use
$\lambda=2$ and a forcing restricted to the first two shells.

The scaling exponents can be defined using the ``correlation
functions" $\left<|u_n|^p\right>$.  It was shown however that the GOY
model suffers from periodic oscillations superposed on the scaling
laws obeyed by these functions \cite{Gat95PRE}. It was proposed in
\cite{Kad95PF} that it is advantageous therefore to focus on the
scaling properties of correlation functions of the type
\begin{equation}
F_q(k_n) \equiv \left<\Big|{\rm Im}\left(u_n u_{n+1}u_{n+2}+{1\over 4} 
u_{n-1}u_{n}u_{n+1}\right)\Big|^{q/3}\right> \ . \label{Fq}
\end{equation}
These functions scale with $k_n$, $F_q(k_n)\sim k_n^{-\zeta_q}$ as
long as $k_n$ is smaller than some dissipative cutoff scale $k_d$ and
sufficiently larger than $k_1$. The exponents
$\zeta_q$ are the scaling exponents whose universality is
questioned. The definition (\ref{Fq}) has the advantage that $\zeta_3$
is known exactly (if one assumes that inertial range scaling is
universal) and it is $\zeta_3=1$ \cite{Piss93PFA}.

The issue to be discussed in this Letter is introduced in Fig. 1 which
shows the results of numerical simulations of this family of
models. The simulations were performed using the adaptive step-size
backward differentiation algorithm (DDEBDF) from the SLATEC library
\cite{DDEBDF,SLATEC}.  We imposed a random forcing on the first two shells, $f_1$= 5
$10^{-3}$ and $f_2=af_1$ where a typical value of $a$ is 0.75. The
time correlation of the forcing has been chosen to be of the magnitude
of the natural forcing turn-over time-scale $\tau= 1/\sqrt{k_1f}=40$.

\narrowtext
\begin{figure}
\epsfxsize=8.5truecm
\epsfbox{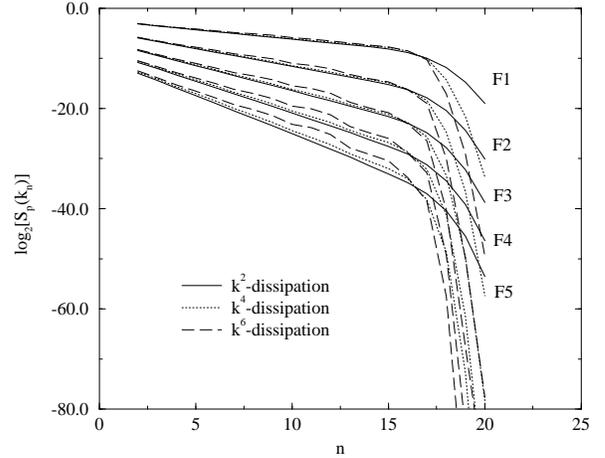}
\caption{The correlation functions $F_1$ to $F_5$ averaged over about 1500 forcing
turn-over scales for a model of 22 shells with a dissipative term
proportional to $k^2$, ($\nu=5\times10^{-7}$), $k^4$, ($\nu=10^{-14}$)
and $k^6$, ($\nu=2\times 10^{-22}$). }
\label{Fig1}
\end{figure}
We plot in  log-log coordinates the correlation functions $F_q$ as a
function of $k_n$, for different values of the parameter
$\alpha=1,2,3$ and $N=22$.  The dissipative cutoff 
\begin{center} \label{N22}
Table~1. Simple scaling regression $k_n^{-\zeta_p}$ 
for regular and hyperviscosity,
$\alpha=1,2,3$. \\~\\
\begin{tabular}{||c|c||c|c|c|c|c||}
\hline 
$k^p$ &N
&$\zeta_1$ &$\zeta_2$ &$\zeta_3$ &$\zeta_4$ &$\zeta_5$ 
\\ \hline  \hline
$k^2$ &22  &0.38 &0.71 &1.00 &1.27 &1.52 \\ \hline
$k^4$ &22  &0.36 &0.69 &0.97 &1.22 &1.46 \\ \hline
$k^6$ &22  &0.34 &0.62 &0.85 &1.06 &1.25 \\ \hline\hline
\end{tabular}
\end{center}

\vskip 0.8cm
\noindent
is chosen (by fixing the parameter $\nu$) to keep $k_d\approx k_0
2^{17}$ for all values of $\alpha$. In Table 1 we show the results
obtained by simple linear fits between the shells 4 and 13.  Results
of this type were interpreted in previous works
\cite{SL95PRL,Schorg95,Dit97PF} as an indication that the inertial
range scaling depends on $\alpha$, casting therefore severe doubts on
the universality of the scaling exponents $\zeta_q$. It should be
stressed that this is not a trivial issue; if indeed the scaling
exponents depend on the dissipative mechanism this could either imply
a major departure from the standard thinking about universality in
turbulence, or an indication that shell models are fundamentally
different from Navier-Stokes dynamics.  The main point of this Letter
is that none of this is necessary. We propose that these scaling plots
should be interpreted as standard corrections to scaling that can be
factored away by taking into acount the effect of the viscous
dissipation. These effect becomes less important when the number of
shells increases and when the length of the inertial range increases,
as we show below in detail.

To provide strong evidence that we are faced with finite size effects 
we integrated the equations of motion for different number of shells. We chose
three sets of data with $N=22$, 28 and 34 shells. In all cases the
integration was performed over 1500 forcing turn-over time scales. The
viscosities were chosen so that the width of the dissipative range is
about 5-6 shells: for $\alpha=1$ (normal dissipation), $\nu=5\times 10^{-7}$,
$8\times 10^{-9}$ and $4\times 10^{-11}$ for $N=22$, 28 and 34
respectively; for $\alpha=2$, $\nu=10^{-14}$, $2\times 10^{-20}$ and
$2\times 10^{-26}$ and for $\alpha=3$, $\nu=2\times 10^{-22}$, $10^{-30}$
and $10^{-41}$. 
The results of this simulation are a direct illustration of the finite size effect: 
for any value
of $\alpha$, we can replot our data for $N=22$, 28 and 34, moving the abscissa
to coalesce the near dissipative region. Consider for example 
$F_3(k_n)$ for which we have a
theoretical expectation, i.e., $F_3(k_n)\sim k_n^{-1}$ in the inertial
range, but for which the data in Table 1 indicate significant deviation
from $\zeta_3=1$. In Fig. 2 we present double-logarithmic plots of
$k_n F_3(k_n)$ 
vs. $k_n$ for $\alpha=2$.  We expect this function to be constant in the inertial
range. Instead, we see a strong viscous effect, leading to deviation
from constancy over 10 shells even before deep dissipation sets
in. Increasing the number of shells moves the problematic region to
higher shells. We have shifted the abscissa in Fig. 2
to collapse all our data together. The data indeed collapses, with an
$N$ independent (fixed size) cross over region.

\begin{figure}\label{k4collapse}
\epsfxsize=8.5truecm
\epsfbox{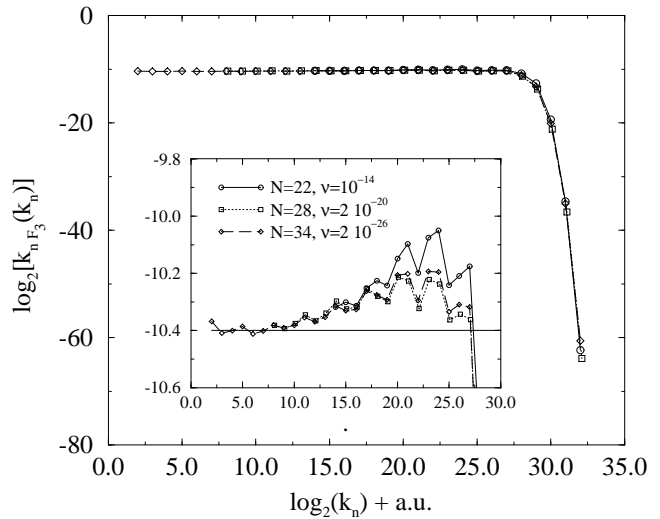}
\caption
{Log-log plots of $k_nS_3(k_n)$ vs. $k_n$ in case of hyperviscosity of
index $\alpha=2$ with different numbers of shells and viscosities.  The
collapse has been obtained by shifting the abscissa. In the inner
frame, we zoom on the y-axis to exhibit the finite size effect
occuring in the inertial range. The solid line shows the constant
behavior expected theroretically. One observes clearly that the
departure from this constant value only occurs in a region of about
ten shells near to the viscous transition. When the inertial range is
large enough, the predicted behavior is recovered.}
\end{figure}
\noindent
This is a sure indication that we are dealing with a finite size
effect and not the breaking of universality. Any deviation of the
apparent $\zeta_3$ from unity is going to disappear slowly when the
size of the inertial range increases. Of course, the linear fit
procedure is very sensitive to the existence of a bump of the type
shown in Fig. 2 and the determination of the exponents in Table 1 is
very unreliable.

As is commonly found in scaling systems that suffer strong finite-size
effects, it is advantageous to fit {\em the whole scaling function}
rather than its power law part alone. In other words, to extract
reliable inertial range scaling exponents, we propose to fit the
correlation functions over their full domain, including the deep
viscous regime. To achieve such a fit, we need to consider first the
viscous range, where we may guess a generalized exponential decay:
 \BE
u_n \sim \exp \left[- \left( \frac{k_n}{k_{d}}\right)^x\right]\ ,
\label{disform} \EE 
where $k_d$ is an appropriate dissipative scale.  In the stationary
state we expect a balance between the non-linar transfer and the
dissipatve damping for every shell. Due to the exponential decay, the
most relevant non-linear term for the $n$'th shell is
$k_{n-1}u_{n-1}u_{n-2}$, which has to balance with $k_n^{2\alpha}
u_n$. Up to logarithmic corrections, the exponent $x$ is then
determined from

\begin{figure}\label{diss}
\epsfxsize=8.5truecm
\epsfbox{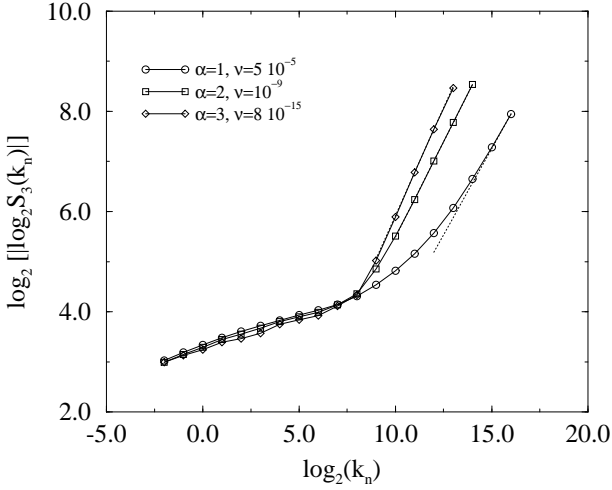}
\caption
{Plots of $\log_2 |\log_2 S_3(k_n)|$ vs $\log_2(k_n)$ for indices of
hyperviscosity $\alpha=1,2,3$ and viscosity $\nu=5\times10^{-5}$,
$10^{-9}$ and $8\times10^{-15}$. The dotted lines represent the
asymptotic stretched exponential behavior. The slopes are respectively
0.69, 0.76 and 0.86 to compare to the theoretical expectation
$x=0.694$.}
\end{figure}
\begin{figure}\label{fitF3}
\epsfxsize=8.5truecm
\epsfbox{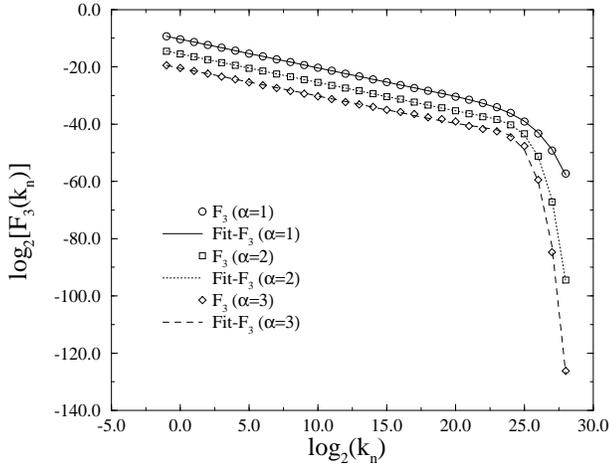}
\caption
{Third order structure function $F_3(k_n)$ for $\alpha=1$, 2 and
3. The symbols indicate the results of simulations with 34 shells;
the lines are the fits performed between the 3rd and
32nd shells. The results have been shifted in the y-coordinate for clarity.}
\end{figure}

\BE
\exp \left[ -\left( \frac{k_{n-1}}{k_{d}}\right)^x  \right] 
\exp \left[ -\left( \frac{k_{n-2}}{k_{d}}\right)^x \right]
\sim \exp \left[ -\left( \frac{k_n}{k_{d}}\right)^x  \right] \ ,
\EE
or equivalently, 
\BE
1+ \lambda^x - \lambda^{2x}=0\quad,
\EE
so that we have finally
\BE
x=\log_\lambda \frac{1+\sqrt{5}}{2}\quad.
\EE

\begin{figure}\label{S3k4k6}
\epsfxsize=8.5truecm
\epsfbox{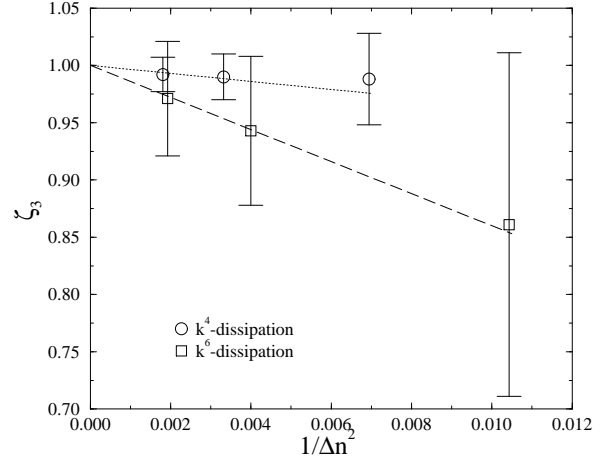}
\caption
{Third order exponent $\zeta_3$ computed via the fitting procedure
described in the text for $\alpha=2$ and 3 obtained for $N=22$, 28,
34. The results have been plotted against $1/(\Delta n)^2$ where
$\Delta n$ is the estimated width of the inertial range. The straight
lines (with respective slope -3.5 and -14) indicate that the
calculated
 exponents
converge linearly toward the predicted value $\zeta_3=1$. }
\end{figure}

\begin{figure}\label{S2k4k6}
\epsfxsize=8.5truecm
\epsfbox{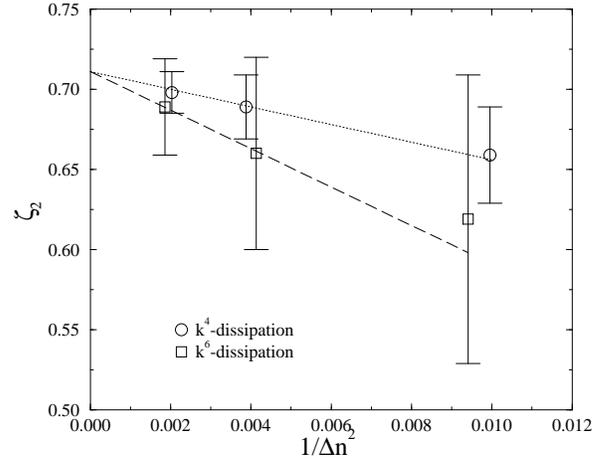}
\caption
{Same figure as above for the second order exponent $\zeta_2$. The
illustrative lines have slopes -6 and -12. The final value
$\zeta_2=0.711$ has been obtained in the normal dissipation case
$\alpha=1$.}
\end{figure}

Our simulations provide excellent support for the deep dissipative
stretched exponential dependence, but the predicted value of $x$
depends on the coefficient $\alpha$, see Fig. 3. In the inertial range
we expect a power law, and to estimate the behavior in the cross over
region we note that the ratio of the viscous to nonlinear term can be
written in the form
\begin{equation}
{\nu k_n^{2\alpha}u_n\over k_n u_n^2}\sim {\nu k_n^
{2\alpha}u_n \over \nu k_d^{2\alpha}u_d}
{k_d u_d^2\over k_n u_n^2}\sim \left({k_n\over k_d}
\right)^{2\alpha-2/3} \ , \label{correct}
\end{equation}
where $u_d$ is the velocity at the viscous cross-over shell with
$k_n=k_d$.  For simplicity we assumed $u_n\sim k_n^{-1/3}$ for this
estimate.  Taking (\ref{correct}) as a first order correction of the
inertial range power law by viscous effects we can propose a
functional form for $F_q(k_n)$ over its full domain:

\BE
F_q(k_n)=A_q k_n^{-\zeta_q} \left(1+\beta_q \frac{k_n}{k_{d,q}}
\right)^{2\alpha-2/3}
\exp\left[-\left(\frac{k_{n}}{k_{d,q}}\right)^{x_q}\right] \ . \label{fit}
\EE
By construction this form fits the scaling law and the deep
dissipative form for $k_n<\ll k_d$ and $k_n\gg k_d$ respectively. For
crossover values of $k_n$ the formula offers two additional free parameters,
i.e. $\beta_q$, $k_{d,q}$. This is the minimal number of fit
parameters that is needed to interpolate between the inertial and the
dissipative ranges. 

As an example of the excellent fit that this formula provides we show
in Fig.4 $F_3(k_n)$ for $\alpha=1$, 2 and 3. This simulation employed
34 shells, and the fit were performed between the 3rd and 32nd shells.
Similarly good fits are found for all the data that we
analyzed. 

Accordingly, we recorded the apparent values of the scaling
exponent $\zeta_q$ for increasing values of the $N$, and noticed that
they converge to a given value. This convergence can be seen in
Figs. 5 and 6 in which the scaling exponents $\zeta_2$ and $\zeta_3$
are plotted for $\alpha=2$, 3 as a function of $1/(\Delta n)^2$ where
$\Delta n=\log_2(k_d/k_2)$ is the estimated width of the inertial
range, and $k_d$ is obtained from the fit. It is obvious that the
exponents converge towards a universal value that agrees with
$\alpha=1$.

The conclusion of this Letter is that it is dangerous to measure
scaling exponents in multiscaling situations without taking into
account crossover and finite size effects. Even in the case of shell
models which appear to have much longer inertial ranges than
Navier-Stokes turbulence, simple log-log plots are misleading. We
propose that in future determination of scaling exponents in numerical
experiments similar care must be given to such issues.

\acknowledgments
This work was supported in part by the European Union under contract
FMRX-CT-96-0010 and the Naftali and Anna Backenroth-Bronicki Fund for
Research in Chaos and Complexity. We thank L. Biferale for
discussions.

\end{multicols}
\end{document}